\documentclass{article}
                                                                           
\usepackage{latexsym, amssymb}

\newcommand{\R}{\mathbb{R}}

\newcommand{\supp}{\mathrm{supp}\,}

\title{Global classical solutions to the spherically symmetric
Nordstr\"om-Vlasov system}
\date{}
\author{H\aa kan Andr\'easson, Simone Calogero\\
        Department of Mathematics, Chalmers University of Technology,\\
        S-41296 G\"oteborg, Sweden\\ 
        Gerhard Rein\\
        Department of Mathematics, University of Bayreuth,\\
        D-95440 Bayreuth, Germany} 
                                                                    
\begin{document}
\maketitle
\begin{abstract}
Classical solutions of the spherically symmetric Nordstr\"{o}m-Vlasov
system are shown to exist globally in time. The main motivation for
investigating the mathematical properties of the Nordstr\"{o}m-Vlasov
system is its
relation to the Einstein-Vlasov system. The former is not a physically
correct model, but it is expected to capture some of the typical
features of the latter, which constitutes a physically satisfactory, 
relativistic model but is mathematically much more complex.  
We show that classical solutions
of the spherically symmetric Nordstr\"{o}m-Vlasov system exist
globally in time for compactly supported initial data under the
additional condition that there is a lower bound on the modulus of the
angular momentum of the initial particle system. We emphasize that
this is not a smallness condition and that our result holds for
arbitrary large initial data satisfying this hypothesis. 

\end{abstract}

\section{Introduction}\label{intro}
\setcounter{equation}{0}
In astrophysics, systems such as galaxies or globular clusters
are often modeled as a large ensemble of particles (stars) which
interact only by the gravitational field which they create
collectively. In such systems collisions among the particles are sufficiently
rare to be neglected. Let $\mathfrak{f}=\mathfrak{f}(t,x,p)\geq 0$ denote 
the density of the particles in phase-space, 
where $t\in\R$ denotes time, $x\in\R^3$ position, and $p\in\R^3$
momentum. This density function satisfies the Vlasov equation---a
continuity equation on phase space---coupled to the field equation(s)
for the gravitational field. 
In the non-relativistic case, i.e., for Newtonian gravity, the field is
governed by Poisson's equation and the resulting system is called the 
Vlasov-Poisson system. 
The initial value problem for this system is by now 
well understood, and satisfactory global existence results have been
obtained, cf.\ \cite{LP,Pf,R,Sch}. In general relativity,
the Poisson equation is substituted by Einstein's equations, which 
coupled to the Vlasov equation yield the Einstein-Vlasov system.  
In contrast to the Vlasov-Poisson system this system is 
highly non-trivial even when matter is left out, since the Einstein
equations by themselves constitute a 
nonlinear system of PDE's. Little 
is known in general about the global structure of vacuum solutions, and
even less is known for matter spacetimes. However,
for small initial data
there is a global existence result for general, asymptotically flat vacuum
spacetimes \cite{CK}. By imposing symmetry 
conditions on spacetime global existence results are known 
also for initial data unrestricted in size. Some of these results also
hold for matter spacetimes when matter is described by 
the Vlasov equation, i.e., for the Einstein-Vlasov system. It should be
pointed out that similar global results for other phenomenological
matter models are not known, indicating that in general relativity
a kinetic description of
matter is mathematically convenient. For a
review on global results for the Einstein equations and for the
Einstein-Vlasov system, cf.\ \cite{Rl,And}. The global existence
results that hold 
for unrestricted data for the Einstein-Vlasov system all
concern cosmological spacetimes which are spatially
compact. In the asymptotically flat case, where for example gravitational
collapse of an isolated body is studied, the global structure of
solutions to the Einstein-Vlasov system is not known 
for large initial data, even in the spherically symmetric case. Some
qualitative information is obtained in \cite{RRS}. 
On the other hand, for small spherically symmetric initial data,
spacetime is known to be geodesically 
complete and thus to contain no 
singularities \cite{RR}. For large data singularities will 
form---cf.\ \cite{Rl2,RRS2}---and 
the central problem of weak cosmic censorship in general relativity
can be studied, i.e., to show that formation of singularities will
always result in black holes. The significance of the latter
problem motivates a study of a less complex but related model, the 
Nordstr\"{o}m-Vlasov system. 

\section{The Nordstr\"{o}m-Vlasov system} 
\setcounter{equation}{0}
In the present paper we investigate a relativistic model 
which is obtained by coupling the Vlasov equation to the Nordstr\"om
scalar gravitation theory \cite{No}. In this theory, the gravitational
effects are mediated by a scalar field $\phi$, and the system 
reads  

\begin{equation} \label{wave1}
\partial_t^2\phi-\bigtriangleup_x\phi=-e^{4\phi}
\int\mathfrak{f}\,\frac{dp}{\sqrt{1+p^{2}}},
\end{equation}
\begin{equation} \label{vlasov1}
\partial_{t}\mathfrak{f} + \widehat{p}\cdot\nabla_x\mathfrak{f} -
\left[\left(\partial_{t}\phi + \widehat{p}\cdot\nabla_x\phi \right) p +
(1+p^{2})^{-1/2}\nabla_x\phi\right]\cdot\nabla_p\mathfrak{f}=0.
\end{equation}
Here  
\[
\widehat{p} :=\frac{p}{\sqrt{1+p^2}}
\]
denotes the relativistic velocity of a particle with momentum $p$,
$p^2 = |p|^2$,
and units are chosen such that the mass of each particle, 
the gravitational constant, and the speed of light are 
all equal to unity.
A solution $(\mathfrak{f},\phi)$ of this system 
is interpreted as follows: The spacetime is a 
Lorentzian manifold with a conformally flat metric which, in the coordinates 
$(t,x)$, takes the form
\[
g_{\mu\nu}=e^{2\phi} \textrm{diag}(-1,1,1,1).
\]
The particle distribution defined on the mass shell
in this metric is given by
\begin{equation} \label{fph}
\mathfrak{f}_\mathrm{ph}(t,x,p)=\mathfrak{f}(t,x,e^\phi p).
\end{equation}
More details on the derivation of this model are given in \cite{Cal}.
It should be emphasized that although the system does not constitute a
physically correct model it still captures some of the essential
features of the Einstein-Vlasov system which are not present in the
Vlasov-Poisson system. In particular, the Nordstr\"om-Vlasov model
does allow for propagation of gravitational waves, even in the
spherically symmetric case. The hope is that the analysis
of this model will lead to a better mathematical understanding
of a whole class of nonlinear partial differential equations
and eventually to a better understanding of the Einstein-Vlasov
system. For previous studies of the Nordstr\"{o}m-Vlasov system we
refer to \cite{CR1,CR2}, where existence of local classical and global 
weak solutions is
established, and to \cite{CL} where the non-relativistic limit is
considered. 

It turns out to be convenient to rewrite the system
in terms of the new unknowns $(f,\phi)$, where $f$ is given by
\begin{equation} \label{ourf}
f(t,x,p)=e^{4\phi(t,x)} \mathfrak{f}(t,x,p).
\end{equation}
The system then takes the form
\begin{equation}\label{wave2}
\partial_t^2\phi-\bigtriangleup_x\phi=-\mu,
\end{equation}
\begin{equation} \label{mudef}
\mu(t,x) = \int f(t,x,p)\,\frac{dp}{\sqrt{1+p^2}},
\end{equation}
\begin{equation} \label{vlasov2}
Sf
- \left[(S\phi)\,p + (1+p^2)^{-1/2} \nabla_x\phi \right]\cdot\nabla_p f
= 4 f\, S\phi\,,
\end{equation}
where
\[
\quad S :=\partial_t+\widehat{p}\cdot\nabla_x
\]
is the free-transport operator.
The function $\mu$
is the trace of the energy momentum tensor.
We supply the system with the initial conditions 
\begin{equation} \label{data}
f(0,x,p)=f^\mathrm{in}(x,p),\quad \phi(0,x)=\phi_0^\mathrm{in}(x),
\quad\partial_t
\phi(0,x)=\phi_1^\mathrm{in}(x),\quad x,\;p \in \R^3, 
\end{equation} 
and we assume that the initial data have the regularity 
\begin{equation} \label{datareg}
f^\mathrm{in} \in C^1_c(\R^6),\ \phi_0^\mathrm{in} \in C^3_b (\R^3),\
\phi_1^\mathrm{in} \in C^2_b (\R^3). 
\end{equation}
Here the subscript $c$ indicates that the functions under consideration
have compact support, $b$ indicates that they are bounded together with
their derivatives up to the indicated order.

We impose spherical 
symmetry on the initial data. By uniqueness, spherical symmetry propagates
from the initial data so that the solution $(f,\phi)$ is also 
spherically symmetric, i.e.,
\[
f(t,A x,A p) = f(t,x,p),\ \phi(t,A x) = \phi(t,x),\ x,p  \in \R^3,\ t \geq 0,\
A \in \mathrm{SO}(3). 
\]
By abuse of notation we can then write
\[
f(t,x,p) = f(t,r,u,\alpha),\ \phi(t,x)= \phi(t,r)
\]
where
\[
r:=|x|,\ u:=|p|,\ \alpha := \angle(x,p),
\]
the latter denoting the angle between $x$ and $p$. In particular,
\[
\nabla_x \phi(t,x) = \partial_r \phi(t,r) \frac{x}{r}.
\] 
Consider the characteristic system
\[
\dot x = \widehat{p},\ \dot p = 
- (S\phi)\,p - (1+p^2)^{-1/2} \nabla_x\phi
\]
of the Vlasov equation (\ref{vlasov2}).
Due to spherical symmetry the angular momentum is conserved
along solutions of the characteristic system. Indeed, defining 
\[
L:= r^2 u^2 \sin^2 \alpha = | x \times p|^2, 
\]
we find that along characteristics
\begin{eqnarray}
&&\frac{d}{ds}
\left(e^{2\phi(s,x(s))} L(x(s),p(s))\right) 
=
2 e^{2\phi} L (\partial_t \phi + \nabla_x \phi \cdot \widehat{p} ) \nonumber\\
&&
\qquad\qquad \ {} +  
2 e^{2\phi} (x \times p)\cdot 
\left[ \widehat{p} \times p - x \times \left((S\phi) p + (1+p^2)^{-1/2}
\partial_r \phi \frac{x}{r}\right)\right] \nonumber\\
&&\qquad\qquad =
2 e^{2\phi} L S\phi - 2 e^{2\phi} L S\phi =
0. \label{angmom}
\end{eqnarray}
The initial datum $f^\mathrm{in}$ is 
assumed to satisfy the additional condition that there 
is a positive number $L_0$ such that 
\begin{equation} \label{dataass}
L(x,p) = r^2 u^2 \sin^2 \alpha \geq L_0 > 0,\ (x,p) \in \supp f^\mathrm{in}.
\end{equation}

In \cite{CR1,CR2} a continuation criterion has been proved for the
Nordstr\"{o}m-Vlasov system which says that a classical 
solution can be extended beyond $t=T$ if the support of the momentum can be
controlled on the time interval $[0,T[.$ More precisely, if the quantity 
\begin{equation} \label{pdef}
P(t)=\sup \{|p|: 0\leq s< t,\ (x,p) \in \supp f(s)\},
\end{equation}
is bounded on the time interval $[0,T[$ then the solution can be
extended to a larger time interval $[0,T+\delta[$ for some
$\delta>0.$ Hence, if $[0,T[$ is assumed to be the maximal life
span of a solution, a bound of $P(t)$ on $[0,T[$
implies that $T=\infty.$ In this work we show that the quantity
$P(t)$ can be controlled if the initial data satisfy the
condition (\ref{dataass}). Our main result can now be formulated.

\smallskip 

\noindent{\bf Theorem} {\em 
For spherically symmetric initial data
$(f^\mathrm{in},\phi^\mathrm{in})$ which satisfy the conditions
(\ref{datareg}) and (\ref{dataass}), there 
exists a unique global 
classical solution $(f,\phi) \in C^1([0,\infty[\times\R^6) 
\times C^{2}([0,\infty[\times\R^3)$ 
of the Nordstr\"om-Vlasov system (\ref{wave2}), (\ref{mudef}),
(\ref{vlasov2}), which satisfies the initial conditions (\ref{data}). 
} 

For the proof it will be essential to look at the behaviour of
another quantity along characteristics (besides the angular momentum):
Along characteristics,
\begin{equation} \label{parten}
\frac{d}{ds} \left(e^\phi \sqrt{1+p^2}\right)
= \partial_t \phi \, e^\phi \frac{1}{\sqrt{1+p^2}}.
\end{equation}
It is of interest to note that this relation is independent of the
symmetry assumption; the quantity under consideration is the particle energy
which would be conserved for a static solution.

\section{Proof of the theorem}
\setcounter{equation}{0} 
We assume that the length of the maximal existence interval
$T$ is finite and show in three steps that the quantity $P$
from (\ref{pdef}) is bounded on $[0,T[$ which then contradicts the
continuation criterion \cite[Prop.~2 ]{CR2}.

\noindent
{\em Step1: An estimate for $\partial_t \phi$}\\
We split $\phi$ as follows:
\begin{equation} \label{asplitphi}
\phi = \phi_\mathrm{hom} + \psi .
\end{equation}
Here $\phi_\mathrm{hom}$ is the solution of the homogeneous wave equation
with the given data, and $\psi$ is the solution of the inhomogeneous
wave equation with zero data. Clearly,
\begin{equation} \label{aphihombound}
\|\phi_\mathrm{hom} (t)\|_\infty
+ \|\partial_t \phi_\mathrm{hom} (t)\|_\infty \leq C,\ 0 \leq t < T;
\end{equation}
recall that we assume $T<\infty$.
By \cite[Lemma 7]{GS2},
\[
\psi (t,x) =
- \int_{|x-y|\leq t}\mu(t-|x-y|,y) \frac{dy}{|x-y|} =
-\frac{1}{2r}\int_0^t\int_{|r-t+\tau|}^{r+t-\tau}\mu(\tau,\lambda)
\,\lambda\,d\lambda\, d\tau.
\]
Note that $\psi\leq 0$ so that $\phi\leq\phi_{\mathrm{hom}}<C.$  
Distinguishing the cases $0<r<t$ and $r\geq t$ we get 
\begin{equation}
\partial_t\psi(t,r) \leq
\frac{1}{2r}\int_0^{(t-r)_+} \mu(\tau,t-r-\tau)\,(t-r-\tau)\,d\tau, 
\label{dtfield}
\end{equation}
where $(t-r)_+$ denotes the positive part of $t-r$.
Define
\[
\widetilde{P}(t) := 
\sup \{e^{\phi(s,x)} |p|\, | (x,p) \in \supp f(s),\ 0\leq s \leq t \}.
\]
Then by conservation of angular momentum,
\begin{eqnarray}
r \mu(t,r)
&=&
2 \pi \int_0^\infty \int_0^\pi f(t,r,u,\alpha) \, r\, \sin\alpha\, u \,d\alpha 
\frac{u}{\sqrt{1+u^2}} du \nonumber \\
&\leq&
C e^{-\phi(t,r)}\int_0^\infty \int_0^\pi f(t,r,u,\alpha)\,d\alpha\,du \nonumber \\
&=&
C e^{3 \phi(t,r)} \int_0^{e^{-\phi(t,r)} \widetilde{P}(t)}
\int_0^\pi  e^{- 4 \phi(t,r)}f(t,r,u,\alpha)\,d\alpha\,du \nonumber \\
&\leq&
C e^{2 \phi(t,r)} \widetilde{P}(t) \leq
C \widetilde{P}(t); \label{armuest}
\end{eqnarray}
recall that $\phi \leq \phi_\mathrm{hom} \leq C$ and
\begin{equation}\label{areprf}
f(t,x,p)=f^\mathrm{in}(X(0),P(0))
\exp \left[4\phi(t,x)-4\phi^\mathrm{in}_0(X(0))\right] \leq C,
\end{equation}
where $(X(s),P(s))$ are the characteristics of the Vlasov equation
with $(X,P)(t) = (x,p)$. If we substitute (\ref{armuest}) into (\ref{dtfield})
we get the estimate
\begin{equation} \label{dtpsiest}
\partial_t \psi(t,r)
\leq 
\frac{C}{r} \widetilde{P}(t),\
r>0,\ 0\leq t <  T.
\end{equation}

\noindent
{\em Step 2: A bound on $\widetilde{P}$}\\
With (\ref{aphihombound}) and (\ref{dtpsiest}) equation (\ref{parten})
implies that
\[
\frac{d}{ds} \left(e^\phi \sqrt{1+p^2}\right)
\leq
C + C \widetilde{P}(s) \frac{1}{r u} 
\leq 
C + C \widetilde{P}(s) \frac{1}{e^{\phi(s,r)}r u \sin \alpha}.
\]
Here we used the fact that $e^\phi\leq C$. 
At this stage we recall the assumption (\ref{dataass}) on the data, 
namely 
\begin{equation} 
L(x,p) = r^2 u^2 \sin^2 \alpha \geq L_0 > 0,\ (x,p) \in \supp f^\mathrm{in}.
\end{equation}
Using the fact that $e^{2\phi} L$ is constant along characteristics we get
\[
\frac{d}{ds} \left(e^\phi \sqrt{1+p^2}\right)
\leq C + C \widetilde{P}(s),
\]
and by Gronwall, $\widetilde{P}$ is bounded.

\noindent
{\em Step 3: Completion of the proof}\\
Using the bound on $\widetilde{P}$ and (\ref{areprf}) we can bound $\mu$,
\[
\mu(t,x) =
\int_{e^{\phi(t,x)} |p| \leq C} f(t,x,p)\frac{1}{\sqrt{1+p^2}} dp 
\leq
C e^{4 \phi(t,x)} \int_{e^{\phi(t,x)} |p| \leq C} dp \leq C,
\]
and hence also $\phi$,
\[
|\phi(t,x)|
\leq
C + C \int_{|x-y| \leq t} \frac{dy}{|x-y|} \leq C.
\]
Hence the bound on $\widetilde{P}$ implies a bound on $|p|$ over
the support of $f$, by the continuation criterion \cite[Prop.~2]{CR2} 
global existence follows, and 
the proof of the theorem is complete. 

{\bf Acknowledgment:}
S.~C.\ acknowledges support by the European HYKE network 
(contract HPRN-CT-2002-00282).


\begin{thebibliography}{99}
\bibitem{And} 
H.~Andr\'easson: 
The Einstein-Vlasov System/Kinetic Theory. 
{\em Living Rev. Relativity }{\bf 5}, (2002) 

\bibitem{Cal} 
S.~Calogero:
Spherically symmetric steady states of 
galactic dynamics in scalar gravity. 
{\em Class.\ Quantum Grav.}\
{\bf 20}, 1729--1741 (2003)

\bibitem{CL}
S.~Calogero, H.~Lee:
The non-relativistic limit of the Nordstr\"om-Vlasov system.
math-ph/0309030 

\bibitem{CR1} 
S.~Calogero, G.~Rein: 
On classical solutions of the Nordstr\"om-Vlasov system.
{\em Commun.\ Partial Diff.\ Eqns.} To appear. 

\bibitem{CR2} 
S.~Calogero, G.~Rein: 
Global weak solutions to the Nordstr\"om-Vlasov system.
math-ph/0309046

\bibitem{CK}
D.~Christodoulou, S.~Klainerman: 
{\em The Global Nonlinear Stability
  of Minkowski Space}. Princeton University Press, 1993. 

\bibitem{GS2}
R.~T.~Glassey, W.~Strauss: 
Absence of Shocks in an initially
dilute collisionless plasma. {\em Commun. Math. Phys. }
{\bf 113}, 191--208 (1987) 

\bibitem{LP}
P.-L.~Lions, B.~Perthame:
Propagation of moments and regularity for the 3-dimensional
Vlasov-Poisson system.
{\em Invent.\ Math.}\ 
{\bf 105}, 415--430 (1991)

\bibitem{No} 
G.~Nordstr\"om: 
Zur Theorie der Gravitation vom Standpunkt des Relativit\"atsprinzips. 
{\em Ann.\ Phys. Lpz.}\ 
{\bf 42}, 533 (1913)  

\bibitem{Pf}
K.~Pfaffelmoser:
Global classical solutions of the Vlasov-Poisson system in three
dimensions for general initial data.
{\em J.\ Diff.\ Eqns.}\
{\bf 95}, 281--303 (1992)

\bibitem{R}
G.~Rein:
Selfgravitating systems in Newtonian theory---the Vlasov-Poisson system.
{\em Banach Center Publications}
{\bf 41}, Part I, 179--194 (1997)

\bibitem{RR}
G.~Rein, A.~D.~Rendall:
Global existence of solutions of the spherically symmetric
Vlasov-Einstein system with small initial data.
{\em Commun.\ Math.\ Phys.}\ 
{\bf 150}, 561--583 (1992)

\bibitem{RRS}
G.~Rein, A.~D.~Rendall, J.~Schaeffer:
A regularity theorem for the spherically symmetric Vlasov-Einstein
system.
{\em Commun.\ Math.\ Phys.}\
{\bf 168}, 467--478 (1995)
 
\bibitem{RRS2}
G.~Rein, A.~D.~Rendall, J.~Schaeffer: 
Critical collapse of collisionless matter: A numerical investigation. 
{\em Phys. Rev. D,}  
{\bf 58}, 044007-1-044007-8, (1998). 

\bibitem{Rl}
A.~D.~Rendall:
Theorems on existence and global dynamics for the Einstein equations. 
{\em Living Rev. Relativity }{\bf 5}, (2002) 

\bibitem{Rl2}
A.~D.~Rendall: 
Cosmic censorship and the Vlasov equation,
{\em Class. Quantum Grav. }{\bf 9}, L99--L104 (1992).

\bibitem{Sch}
J.~Schaeffer:
Global existence of smooth solutions to the Vlasov-Poisson system
in three dimensions.
{\em Commun.\ Part.\ Diff.\ Eqns.}\ 
{\bf 16}, 1313--1335 (1991)


\end{thebibliography}
\end{document}